\documentclass[prl,twocolumn,showpacs,nofootinbib]{revtex4}
\usepackage{amsfonts,amsmath}
\usepackage{graphicx}
\usepackage{color}

\begin{document}
\title{Superballistic wavepacket spreading in double kicked rotors}
\author{Ping Fang and Jiao Wang}
\email{phywangj@xmu.edu.cn}
\affiliation{Department of Physics and Institute of Theoretical Physics
and Astrophysics, Xiamen University, Xiamen 361005, Fujian, China}

\begin{abstract}
We investigate possible ways in which a quantum wavepacket spreads. We
show that in a general class of double kicked rotor systems, a wavepacket
may undergo superballistic spreading; i.e., its variance increases as the
cubic of time. The conditions for the observed superballistic spreading
and two related characteristic time scales are studied. Our results 
suggest that the symmetry of the studied model and whether it is a 
Kolmogorov-Arnold-Moser system are crucial to its wavepacket spreading
behavior. Our study also sheds new light on the exponential wavepacket
spreading phenomenon previously observed in the double kicked rotor
systems.
\end{abstract}

\pacs{05.45.Mt, 05.60.Gg, 03.75.-b}

%\PACS number(s): 05.45.Mt, 05.60.Gg, 03.75.-b
%05.45.Mt Quantum chaos; semiclassical methods
%05.60.Gg Quantum transport
%03.75.-b Matter waves

\date{\today }
\maketitle

\section{I. Introduction}

In a quantum system, a wavepacket usually spreads following a power law
of time, i.e., its variance increases in time as $\sim t^\gamma$, with
$\gamma$ being a constant and $0 \le \gamma \le 2$. For $\gamma=0$, the
wavepacket will be localized, while for $\gamma=2$, the wavepacket will
spread ballistically. The case of $\gamma=1$ is referred to as `normal
diffusion', in contrast to two `anomalous diffusion' cases, i.e.,
subdiffusion for $0<\gamma<1$ and superdiffusion for $1<\gamma<2$. It
is found that under certain conditions the exponent $\gamma$ can be
related to the fractal dimension of the system's spectrum~\cite{Geisel,
Italo89, Italo93}. For example, for the two special cases of localization
($\gamma=0$) and ballistic spreading ($\gamma=2$), the spectrum is
discrete and absolutely continuous, respectively.

In recent years, investigations of the possible ways in which a quantum
wavepacket spreads have led to some important findings. An interesting
example is that~\cite{Kottos}, if a segment of a one-dimensional (1D)
homogenous lattice is replaced by a segment of disordered structure, then
a wavepacket that initially resides on the implanted segment may spread
`superballistically' with $2<\gamma\le 3$, until a certain characteristic
time $T$ that depends on the length of the implanted segment: The
longer the latter is, the longer $T$ is. When the time
exceeds $T$, the wavepacket tends to converge asymptotically to the
ballistic spreading. In a more recent study~\cite{Tong}, it was found
that for a 1D tight-binding lattice without on-site potential, if one
implants a segment of lattice with on-site potential, then a wavepacket
initially prepared on the latter may not only spread superballistically,
but also hyperdiffusively, i.e., $\gamma$ can be as large as $3<\gamma<5$.

More interestingly, power laws are not the only ways in which a wavepacket
spreads. It has been found that a wavepacket can spread in time even
exponentially~\cite{Wang11}. Similar to the two cases~\cite{Kottos, Tong}
mentioned above, the time for which the exponential spreading lasts
depends on the system's parameters, which is finite but in principle can
be infinite as the system's parameters are tuned. This finding unveils
a new type of quantum motion.

The model system in which the exponential spreading was found is a
variant of the quantum kicked rotor (QKR)~\cite{Casati79}. Despite its
seeming simplicity in constructure, the QKR exhibits very rich dynamics
and has played a central role in quantum chaos studies. Up to now a wide
spectrum of wavepacket spreading ways, from power laws with $0\le
\gamma \le 2$ to exponential laws, has been found in the QKR as well as
its variants, but superballistic and hyperdiffusive spreading for $\gamma
\ge 3$ has not been reported yet.

In this work we show that the superballistic spreading with $\gamma=3$
is also possible in a class of kicked rotor systems. This result implies
that the superballistic spreading exists in very general systems, not
only restricted in the lattices with hybrid structures~\cite{Kottos,
Tong}. In addition, it suggests a new type of quantum motion in the
QKR, evidencing again the dynamics wealth of this paradigmatic quantum
chaos model. As various QKR systems have been realized with atom-optics
setup~\cite{Moore, Chabe}, our result also suggests a possible way to
experimental studies of the superballistic spreading.

Based on our understanding of the mechanism for the exponential
spreading~\cite{Wang11, Wang13}, our strategy here is to design variant
QKR systems such that the superballistic diffusion can happen in the
pseudoclassical limit~\cite{Rebuzzini}. Then we show and study the
superballistic wavepacket spreading in the original systems, in particular
its conditions and characteristics. This method involves the concepts of
quantum resonance~\cite{QR80}, quantum antiresonance~\cite{QR80, Dana96},
pseudoclassical limit theory~\cite{Rebuzzini}, Kolmogorov-Arnold-Moser
(KAM) systems~\cite{KAM}, etc. For recent progress in
revealing the general wave propagation properties in quantum mechanics
and the roles nonlinearities may play in matter wavepacket dynamics, we
refer the reader to Refs.~\cite{WangPX, Shapira, Sui}. In the following
we will first describe the models to be focused on in Sec.~II, then show
their quantum wavepacket spreading in Sec.~III and discuss their
pseudoclassical systems, the mechanism and the rate of the observed
superballistic spreading in Sec.~IV. Two characteristic time scales
will be analyzed in Sec.~V. Finally, we will make extended discussions
and conclude in Sec. VI.

\section{II. Models}

The 1D kicked rotor system~\cite{Chirikov} is composed of a point
particle confined to move on a circle. The motion of the particle is
subjected to a series of sudden kicks imposed periodically, otherwise
its motion on the circle is free. Assuming that the inertial moment of
the particle and the radius of the circle are unitary, the Hamiltonian
of the system is
\begin{equation}
H=\frac{p^2}{2}+KV(\theta)\sum_{n\in\mathbb{Z}}\delta(t-n\tau).
\end{equation}
Here $p$ and $\theta\in[-\pi,\pi)$ are the angular momentum and the
angular position of the particle, respectively, $V(\theta)$ is the
external field of period $2\pi$, i.e., $V(\theta)=V(\theta+2\pi)$, and
$K$ is a parameter that controls the amplitude of the kicks turned on
instantaneously at multiples of $\tau$. In the standard kicked rotor model
the potential is $V_S (\theta)=\cos(\theta)$, which is both nondegenerate
and sufficiently smooth. Hence the classical standard kicked rotor is
a KAM system~\cite{KAM} so that for $K\tau<\kappa_C$ (where $\kappa_C
\approx 1$ is a critical value) the motion of the system in the angular
momentum space is confined by the Cantori and is localized, but for
$K\tau> \kappa_C$ the motion becomes unbounded due to the breaking of
all the Cantori, and on average its energy increases linearly in time
\cite{Chirikov}, i.e, $E(t)=p^2/2 \sim K^2 t$. In sharp contrast, the
motion of the quantum standard kicked rotor depends on wether $\hbar\tau$
is an irrational multiple of $\pi$ (here $\hbar$ is the effective Planck
constant of the system). If yes, then the energy of the system will
eventually saturate, known as the dynamical localization~\cite{Casati79,
Chirikov}. Otherwise the so-called quantum resonance occurs, i.e., the
energy increases quadratically in time, unless $\hbar\tau$ is an odd
integer multiple of $2\pi$, at which the state of the system goes back
to itself after every two kicks and the energy keeps to oscillate
periodically. This special case is known as the quantum
antiresonance~\cite{QR80, Dana96}.

An important variant of the kicked rotor is the double kicked rotor
which was first proposed by Li et al~\cite{DKRChen} to study the chaos
controlling problem in Hamiltonian systems. In this model, during a time
of $\tau$ the rotor is kicked twice separated by a time interval $\Delta
\tau<\tau$. The Hamiltonian is
\begin{equation}
H=\frac{p^2}{2}+KV(\theta)\sum_{n\in\mathbb{Z}}\delta(t-n\tau)
+KV(\theta)\sum_{n\in\mathbb{Z}}\delta(t-n\tau+\Delta\tau).
\end{equation}
Note that the two kicks can be different in general, but in this work we
assume they are the same. Quantum mechanically, if we denote the state
of the system at the time just before $t=n\tau$ as $|\psi(n)\rangle$, and
after a time $\tau$ it evolves into $|\psi(n+1)\rangle$, then the latter
can be obtained from the former as $|\psi(n+1)\rangle=U|\psi(n)\rangle$,
with the evolution operator
\begin{equation}
U=e^{-i(\tau-\Delta\tau)\frac{p^2}{2\hbar}}
  e^{-i\frac{K}{\hbar}V(\theta)}
  e^{-i\Delta\tau\frac{p^2}{2\hbar}}
  e^{-i\frac{K}{\hbar}V(\theta)}.
\end{equation}
The second and the forth factors result from kicks and the first
and the third term represent the free motion between kicks. In general,
the quantum double kicked rotor (QDKR) with potential $V_S$ exhibits
dynamical localization: The wavepacket eventually displays exponentially
decaying tails in the angular momentum space and the localization length
depends on $\hbar$ in a power law with a fractional number exponent~\cite
{DKRexp, Wang07}. The most striking phenomenon of the QDKR appear when
the main quantum resonance condition, i.e., $\hbar\tau=4\pi$, is
satisfied, under which the evolution operator reduces to the following
more symmetric form:
%%%%%%%%%%%%%%%%%%%%%%%%%%%%%%%%%%%%%%%%%%%%%%%%%%%%%%%%%%%%%%%%%%%%%%%%%%%%%%%%%%%%%Eq U
\begin{equation}
U=e^{i\frac{p^2}{2\hbar}}e^{-i\frac{K}{\hbar}V(\theta)}
  e^{-i\frac{p^2}{2\hbar}}e^{-i\frac{K}{\hbar}V(\theta)}.
\label{Eq:U}
\end{equation}
%%%%%%%%%%%%%%%%%%%%%%%%%%%%%%%%%%%%%%%%%%%%%%%%%%%%%%%%%%%%%%%%%%%%%%%%%%%%%%%%%%%%%Eq U
(Without loss of generality we set $\Delta\tau=1$.) This is apparent in
view of the fact that $\exp(-i\tau\frac{p^2}{2\hbar})|j\rangle=|j\rangle$
for $\hbar\tau=4\pi$, where $\{|j\rangle\}$ represent the eigenstates of
the angular momentum $p$, i.e.,
\begin{equation}
p|j\rangle=j\hbar|j\rangle, ~~~\langle\theta|j\rangle=\frac{1}
{\sqrt{2\pi}}e^{ij\theta}, ~~~ j\in \mathbb{Z}.
\end{equation}

%%%%%%%%%%%%%%%%%%%%%%%%%%%%%%%%%%%%%%%%%%%%%%%%%%%%%%%%%%%%%%%%%%%%%%%%%%%%%%%%%%figure1
\begin{figure*}[!]
\includegraphics[width=17.5cm]{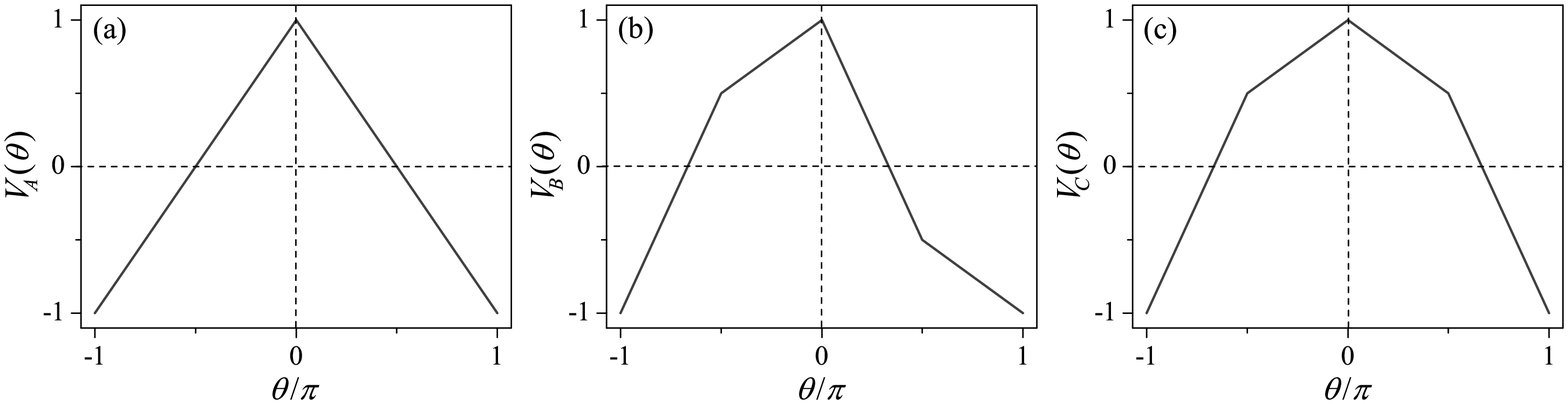}
\caption{Three linear piecewise potential functions of the 1D double
kicked rotor studied in this work. Potential $V_B$ and $V_C$ have the
symmetry of $V_B(\theta)=-V_B(\theta+\pi)$ and $V_C(\theta)=V_C(-\theta)$,
respectively, while potential $V_A$ have both.}
\label{F:V}
\end{figure*}
%%%%%%%%%%%%%%%%%%%%%%%%%%%%%%%%%%%%%%%%%%%%%%%%%%%%%%%%%%%%%%%%%%%%%%%%%%%%%%%%%%figure1

In the following, we will restrict ourselves on the QDKR defined by
the evolution operator given by Eq.~(\ref{Eq:U}). It has been found
that when potential $V_S$ is taken, this system  has the same spectrum
as the kicked Harper model due to the additional symmetry~\cite{Wang08,
Lawton}; Moreover, if $\hbar$ satisfies further the condition $\hbar
\approx 2\pi M/N$, where $M$ and $N$ are coprime odd integers\footnote{
As $\hbar\tau=4\pi$, $\Delta \tau=1$ and $\tau>\Delta \tau$,
the condition $\hbar \approx 2\pi M/N$ can be satisfied by adjusting the
parameter $\tau$ so that $\tau\approx 2N/M>1$. This implies that $2N>M$,
which is assumed in the following.}, 
a quantumwavepacket that can be readily prepared initially may spread in 
time exponentially~\cite{Wang11, Wang13}. As we will show in the following,
the exponential spreading is possible thanks to two properties of $V_S$,
i.e., its KAM nature and the symmetry of $V_S (\theta)=-V_S(\theta+\pi)$.
This potential also has the reflecting symmetry of $V_S(\theta)=V_S(-\theta)$
but this symmetry is irrelevant to the exponential spreading. In order to
show this and to explore the superballistic spreading, in this work we
consider the QDKR [defined by Eq.~(\ref{Eq:U})] with potentials of non-KAM
nature and various symmetries. For simulations, we consider three linear
piecewise potential functions, denoted as $V_A$, $V_B$, and $V_C$, that
are schematically plotted in Fig.~\ref{F:V}. All of them are non-analytic
and therefore non-KAM. In particular, $V_A$ has two symmetries as $V_S$
does, i.e., $V_A(\theta) =-V_A(\theta+\pi)$ and $V_A(\theta)=V_A(-\theta)$,
but $V_B$ and $V_C$ have only the former and the latter, respectively. Our
aim is to compare the wavepacket dynamics of the QDKR for these potentials
to figure out the key factors for the superballistic and exponential
spreading. Our study has clarified that the details of the potential are
irrelevant, but the simulation results presented in the following are for
the three concrete potential functions with common turning points of
$(\theta, p)=(-\pi,-1)$ and $(0,1)$, and extra turning points of $(\theta,
p)=(-\pi/2,g)$ and $(\pi/2,-g)$ for $V_B$ but $(\theta,p)=(-\pi/2,g)$ and
$(\pi/2, g)$ instead for $V_C$. Here $g$ is a parameter that we assume to
be $g=0.5$; if $g=0$ then both $V_B$ and $V_C$ reduce to $V_A$. Throughout
the paper the kicking strength parameter is fixed to be $K=5$ at which
the classical limit of the three systems are all chaotic.

\section{III. Quantum superballistic wavepacket spreading}

For a pure state of the QDKR, the kinetic energy of the
system $E(t)=\langle \psi(t)|\frac{p^2}{2} |\psi(t)\rangle$ also
represents the variance of the wavepacket in the angular momentum space.
We therefore investigate the time dependence of the kinetic energy of
our QDKR models. The initial state is set to be $|\psi(0) \rangle=
|0\rangle$, i.e., the eigenstate of $p$ with a zero angular momentum.
Numerically, we invoke the fast Fourier transform algorithm to
simulate the evolution of the system and to calculate $E(t)$ with
the obtained $|\psi(t)\rangle$.

\begin{figure*}[!]
\includegraphics[width=17.5cm]{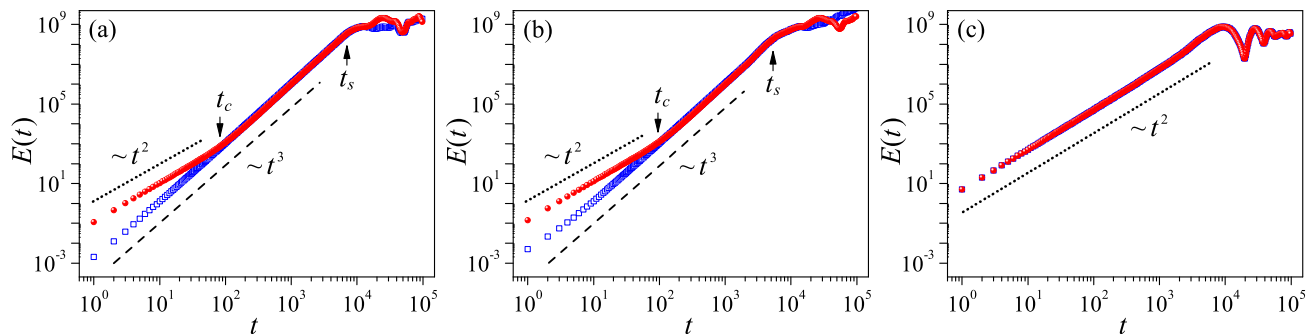}
\caption{Time dependence of the energy of both the QDKR (red solid bullets)
and its pseudoclassical system (blue open squares) for three potential functions,
$V_A$ (a), $V_B$ (b), and $V_C$ (c), respectively. In each panel, the time
scaling $\sim t^2$ and $\sim t^3$ are indicated by the dotted and the dashed
line, respectively, for reference. In all the cases $\hbar=2\pi+\tilde\hbar$
and $\tilde\hbar=10^{-3}$.}
\label{F:E}
\end{figure*}

We have found that if $\hbar$ is close to a value of $2\pi M/N$, where
$M$ and $N$ are odd coprime integers, for the QDKR with potential $V_A$
and $V_B$ the wavepacket undergoes the superballistic spreading. This
is the main result of this work. Examples for the case of $\hbar\approx
2\pi$ and $\hbar\approx 2\pi/3$ are shown in Fig.~\ref{F:E}(a)-(b) and
Fig.~\ref{F:tcts}(c), respectively. Generally, $E(t)$ displays three stages,
and the two characteristic time scales are denoted as $t_c$ and $t_s$
[indicated by the arrows in Fig.~\ref{F:E}(a)-(b)], respectively. At the
first stage for $t<t_c$, the wavepacket spreads ballistically, $E(t)\sim
t^2$; At the second stage for $t_c<t<t_s$, it spreads superballistically,
$E(t)\sim t^3$. Finally, for $t>t_s$, the superballistic spreading is
suppressed and $E(t)$ begins to oscillate around a certain value denoted
as $E_s$. For potential $V_C$ the intermediate superballistic
stage is missing; there are only two stages left: the ballistic stage is
followed by the oscillating one.

%%%%%%%%%%%%%%%%%%%%%%%%%%%%%%%%%%%%%%%%%%%%%%%%%%%%%%%%%%%%%%%%%%%%%%%%%%%%%%%%%%figure2
%%%%%%%%%%%%%%%%%%%%%%%%%%%%%%%%%%%%%%%%%%%%%%%%%%%%%%%%%%%%%%%%%%%%%%%%%%%%%%%%%%figure2

Denote the deviation of $\hbar$ from $2\pi M/N$ as $\tilde\hbar$, i.e.,
$\hbar=2\pi M/N+\tilde\hbar$, our numerical analysis has established
the following scaling\footnote{We assume $\tilde\hbar\ge
0$ throughout but $\tilde\hbar$ can be negative (see Ref.~\cite{Rebuzzini})
and all the results presented in this work can be extended to $\tilde
\hbar < 0$ straightforwardly.}:
\begin{equation}
t_c\sim 1/\sqrt{\tilde\hbar},
~~t_s\sim 1/\tilde\hbar,
~~E_s\sim 1/\tilde\hbar^2.
\label{Eq:tctsEs}
\end{equation}
It implies that though the superballistic stage is finite, it can be
arbitrarily long by decreasing the parameter $\tilde\hbar$. In the
following sections we will discuss further these relations and why
for potential $V_C$ the superballistic spreading does not occur.

\section{IV. Pseudoclassical limit at $\hbar\approx 2\pi$}

For both the QKR and the DQKR with potential $V_S$, the special
case of $\hbar=2\pi+\tilde\hbar$ ($|\tilde\hbar|\ll 1$) is of special
interest because it allows one to analyze the system by a classical
method~\cite{Rebuzzini, Wang11}. This is known as the pseudoclassical
limit theory~\cite{Rebuzzini}, which maps the system onto a virtual
classical system by assuming $\tilde\hbar$ as a virtual Planck constant
and taking the virtual classical limit of $\tilde\hbar \to 0$. We find
that the pseudoclassical limit theory is also valid in dealing with our
DQKR models of non-KAM potentials. In this section we use this method
to explore the mechanism and the properties of the superballistic
spreading. Let
\begin{equation}
\tilde\theta=\theta,
~~\tilde p=p\tilde\hbar/\hbar,
~~\tilde K=K\tilde\hbar/\hbar,
\end{equation}
the operator $U$ defined by Eq.~(\ref{Eq:U}) can be rewritten as
\begin{equation}
\tilde U=
  e^{\frac{i}{\tilde\hbar}(\frac{\tilde p^2}{2}+\pi\tilde p)}
  e^{-i\frac{\tilde K}{\tilde\hbar}V(\tilde\theta)}
  e^{-\frac{i}{\tilde\hbar}(\frac{\tilde p^2}{2}+\pi\tilde p)}
  e^{-i\frac{\tilde K}{\tilde\hbar}V(\tilde\theta)}.
\end{equation}
This implies a rotor that is described by the conjugate pair
$(\tilde\theta, \tilde p)$ and is subjected to the external field
$\tilde K V(\tilde\theta)$. In the pseudoclassical limit of $\tilde
\hbar\to 0$, this rotor undergoes a free rotation with the angular
velocity $d\tilde \theta/dt=\tilde p+\pi$ or $-(\tilde p+\pi)$
alternatively between two consequential kicks, and the corresponding
classical motion within a single step is given by the map $M: (\tilde
\theta_n, \tilde p_n) \to (\tilde\theta_{n+1}, \tilde p_{n+1})$,
\begin{equation}
M: \left\{\begin{array}{ll}
\rho=\tilde p_n+\tilde K f(\tilde\theta_n),\\
o=\tilde\theta_n+\rho+\pi,\\
\tilde p_{n+1}=\rho+\tilde K f(o),\\
\tilde\theta_{n+1}=o-\tilde p_{n+1}+\pi,\\
\end{array}\right.
\label{Eq:M}
\end{equation}
where $f(\theta)\equiv -dV(\theta)/d\theta$ ($\rho$ and $o$ are two
intermediate variables). This defines the classical kicked rotor system
in the pseudoclassical limit, which, in the following we refer to as the
pseudoclassical system of the QDKR. We are interested in the `energy'
of this pseudoclassical system $\tilde E\equiv \langle\tilde p^2/2
\rangle$, where the average is taken over an ensemble of initial
conditions. In order to compare the results with those of the QDKR, we
assume the rescaled energy, denoted by $E(t)$ as well without confusion,
that $E(t)\equiv\tilde E(t)\hbar^2/\tilde\hbar^2$. We also assume the
initial conditions $(\tilde\theta_0,\tilde p_0)$ that match the initial
quantum state $|0\rangle$; i.e., $\tilde p_0=0$ and $\tilde \theta_0$
distributes uniformly in $[-\pi,\pi)$. In our simulations, an ensemble
of $10^4$ initial conditions is adopted.

The results of $E(t)$ for the pseudoclassical systems with the three
linear piecewise potentials are shown in Fig.~\ref{F:E} for $\hbar=2\pi+
\tilde\hbar$ and $\tilde\hbar=10^{-3}$. It can be seen that for potentials
$V_A$ and $V_B$, $E(t)$ increases with the cubic of time all the way until
$t=t_s$. In addition, the agreement between the QDKR systems and their
pseudoclassical systems is perfect for $t>t_c$. For potential $V_C$, the
pseudoclassical system also undergoes ballistic motion, and the agreement
with the QDKR is perfect as well. These results suggest that the
pseudoclassical limit theory works well for our models despite of their
non-KAM nature, and the ballistic spreading of the QDKR with $V_A$ and
$V_B$ is a pure quantum effect.

%%%%%%%%%%%%%%%%%%%%%%%%%%%%%%%%%%%%%%%%%%%%%%%%%%%%%%%%%%%%%%%%%%%%%%%%%%%%%%%%%%figure3
\begin{figure*}[!]
\includegraphics[width=17.5cm]{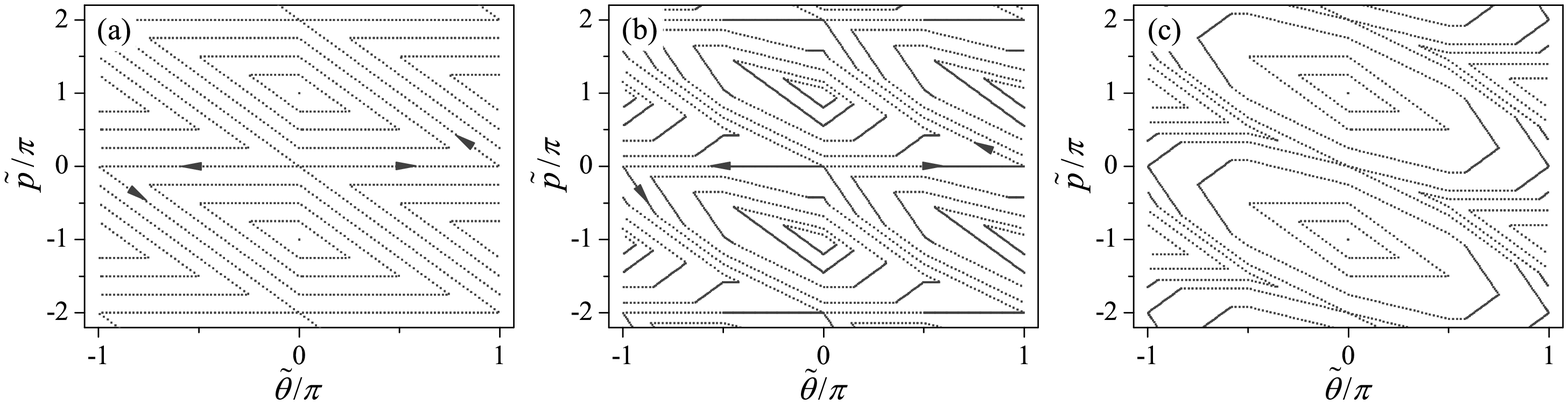}
\caption{The phase space structure of the system in the pseudoclassical
limit of the QDKR with potential given by $V_A$ (a), $V_B$ (b), and
$V_C$ (c), respectively. The arrows in (a) and (b) indicate the moving
direction of phase points on the negative and positive half of $\tilde
\theta$-axis. All the parameters are the same as in Fig.~\ref{F:E}.}
\label{F:PP}
\end{figure*}
%%%%%%%%%%%%%%%%%%%%%%%%%%%%%%%%%%%%%%%%%%%%%%%%%%%%%%%%%%%%%%%%%%%%%%%%%%%%%%%%%%figure3

In order to understand the mechanism of the observed superballistic
(for $V_A$ and $V_B$) and ballistic (for $V_C$) motions, we plot the
phase portraits of the three pseudoclassical systems in Fig.~\ref{F:PP}.
It shows that for $V_A$ and $V_B$ there is a common feature that is
absent for $V_C$: there is a horizontal phase line along the $\tilde
\theta$ axis. For $V_A$ [see Fig.~\ref{F:PP}(a)], checking further the
motion of the phase points one may find that a phase point on the positive
half of $\tilde\theta$-axis moves to the right at a constant speed until
it reaches $\tilde\theta=\pi$, then the point moves up towards the point
$(\tilde\theta,\tilde p)=(0,2\pi)$, also at a constant speed. The motion
of the points on the negative half of $\tilde\theta$-axis is the same but
along the opposite direction. This observation explains why $\tilde E(t)
\sim t^3$. In fact, $\tilde E(t)\sim P_{\tilde p\ne 0}(t)t^2$, where
$P_{\tilde p\ne 0}(t)$ stands for the potion of phase points in the average
ensemble that have left $\tilde\theta$-axis at time $t$, which increases
as $\sim t$, and $t^2$ stands for the contributions of these points to the
averaged energy because their momentum increases linearly; i.e., $\tilde
p(t)\sim t$. This mechanism is the same in spirit as that outlined in
Ref.~\cite{Kottos}. For $V_B$ this explanation still works, despite of
the fact that the motion of phase points on $\tilde\theta$-axis takes
two different constant speeds instead because $|\partial V_B(\theta)/
\partial \theta|$ has two different values. For $V_C$ after one step
of evolution all the phase points in the average ensemble will leave
$\tilde \theta$-axis and begin to move ballistically, hence we have
$P_{\tilde p\ne 0}(t)=1$ for $t>0$ and $\tilde E(t)\sim t^2$.

It is important to note that the phase space structure has a period $2\pi$
in $\tilde p$ direction. Consequently, the superballistic (for $V_A$ and
$V_B$) and the ballistic (for $V_C$) diffusion does not last forever. When
the phase points reach $|\tilde p|\approx 2\pi$ their motion directions may
change, and therefore the energy increasing slows down. For the cases shown
in Fig.~\ref{F:E}, $E_s$ can then be estimated to be $\tilde E_s \hbar^2/
\tilde\hbar^2\approx 8\times10^8$, which agrees very well with the
simulation results (see Fig.~\ref{F:E}).

With the help of the classical motion equation given by map $M$,
we can further determine the prefactor of the superballistic and the
ballistic diffusion. As an example here we consider the case with
potential $V_A$, but the calculations can be extended to other two
cases straightforwardly. Based on Eq.~(\ref{Eq:M}), after one step of
iteration a given initial condition $(\tilde\theta_0,\tilde p_0)$
$(\tilde p_0=0)$ is mapped to
\begin{equation}
(\tilde\theta_{1},\tilde p_{1})=\left\{
\begin{array}{ll}
(\tilde\theta_0-\Delta,2\Delta)&
      {\rm for}~~\pi-\Delta\le\tilde\theta_0<\pi,\\
(\tilde\theta_0+\Delta,0)&
      {\rm for}~~0\le\tilde\theta_0<\pi-\Delta,\\
(\tilde\theta_0-\Delta,0)&
      {\rm for}~~\Delta-\pi\le\tilde\theta_0<0,\\
(\tilde\theta_0+\Delta,2\Delta)&
      {\rm for}~-\pi\le\tilde\theta_0<\Delta-\pi,\\
\end{array}\right.
\end{equation}
where $\Delta\equiv 2\tilde K/\pi$. It shows that after one step
of iteration, a portion of $\Delta/\pi$ phase points in the initial
condition ensemble increases by an amount of $2\Delta^2$ in their energy.
Repeating this calculation, one finds that after $n$ steps of iteration,
$n$ portions of initial conditions have, respectively, energy $2\Delta^2$,
$4\Delta^2$, $\cdots$, $2n^2 \Delta^2$, before the momentum of the portion
with the largest energy $2n^2 \Delta^2$ reaches $|\tilde p| \approx 2\pi$.
It follows that in the time range $t<t_s$ with
\begin{equation}
t_s\approx \frac{\pi}{\Delta}=\frac{\pi^2\hbar}{2K\tilde\hbar}
\label{Eq:ts}
\end{equation}
the ensemble averaged energy increases as
\begin{equation}
\tilde E(t)=\frac{\Delta^3}{3\pi}t(t+1)(2t+1)\approx\frac{2\Delta^3}
{3\pi}{t^3},
\end{equation}
or in terms of $E(t)$, that
\begin{equation}
E(t)=\tilde E(t)\frac{\hbar^2}{\tilde\hbar^2}\approx\frac{16K^3\tilde\hbar}
{3\pi^4\hbar}{t^3}.
\label{Eq:t3ratio}
\end{equation}
These results have been fully corroborated by the simulations.

Now it is in order to explain why for $V_A$ and $V_B$ there is the crucial
phase space structure along $\tilde\theta$-axis but for $V_C$ there is not.
For an initial phase point $(\tilde \theta_0, \tilde p_0)$ on $\tilde\theta$-axis,
i.e., $\tilde p_0=0$, it can be obtained from Eqs.~(9) that $\tilde p_1=\tilde
K[f(\tilde \theta_0)+f(\tilde \theta_0+\tilde K f(\tilde\theta_0)+\pi)]$;
Considering that $\tilde K=\tilde\hbar K/\hbar$ can be much smaller than 
one due to $\tilde\hbar\ll 1$, it implies that in general $\tilde p_1\approx 
\tilde K [f(\tilde \theta_0)+f(\tilde \theta_0+\pi)]$. So after one step of 
iteration a typical initial phase point will leave $\tilde\theta$-axis. However, 
if the potential has the symmetry $V(\tilde \theta)=-V(\tilde \theta+\pi)$, 
then $f(\tilde\theta)$ has the same symmetry and as a result $\tilde p_1=\tilde 
K[f(\tilde \theta_0)-f(\tilde\theta_0+\tilde K f(\tilde\theta_0))]=O(\tilde K^2)$ 
if $V$ is smooth enough. Therefore, up to the first order of $\tilde K$ we have 
$\tilde p_1=0$. (In fact, for $V_A$ and $V_B$, we can prove that $p_1=0$ exactly
if $\tilde\theta_0$ and $\tilde\theta_0+\tilde K f(\tilde\theta_0)$ are on the 
same line segment of the potential function.) This argument works not only 
for $V_A$ and $V_B$, but also for other non-KAM potentials (see one example 
presented in Fig.~5) and KAM potentials (such as $V_S$) of the same symmetry. 
For a KAM system, the difference is that there are hyperbolic fixed points 
on $\tilde\theta$-axis; they make the nearby phase points approach and leave 
them exponentially and thus induce the exponential wavepacket spreading 
instead~\cite{Wang11, Wang13}.

\section{V. Two characteristic time scales}

Now let us come back to the QDKR models. As their pseudoclassical systems
mimic their motions closely, we are able to explore their motions via the
latter. In this section we discuss the times $t_c$ and $t_s$ that
characterize the quantum superballistic spreading. For the sake of
convenience let us first consider the case of potential $V_A$ and $\hbar=
2\pi+\tilde\hbar$ ($\tilde\hbar\ll 1$). For this case we have obtained
$t_s$ in the pseudoclassical system and found that it applies to the QDKR
system equally. However, as the pseudoclassical system does not show the
ballistic motion in the initial stage for $t<t_c$, we can not probe the
clues of the time $t_c$ from it.

%%%%%%%%%%%%%%%%%%%%%%%%%%%%%%%%%%%%%%%%%%%%%%%%%%%%%%%%%%%%%%%%%%%%%%%%%%%%%%%%%%figure4
\begin{figure*}[!]
\includegraphics[width=17.5cm]{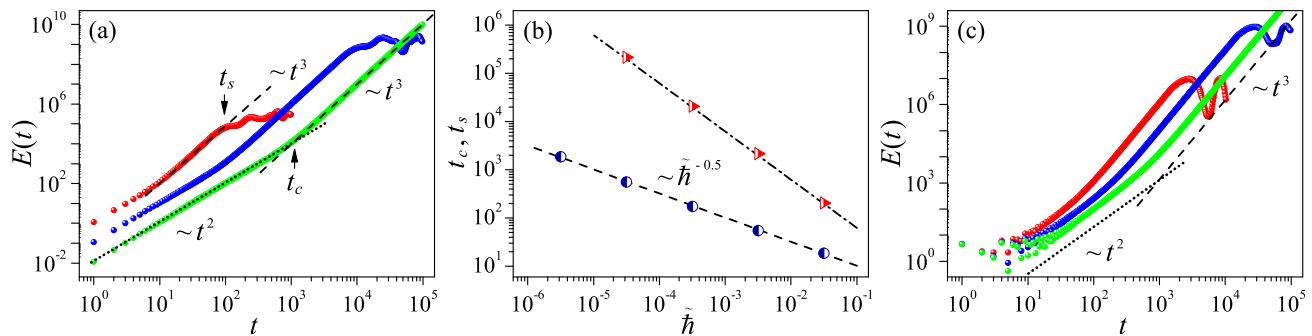}
\caption{(a) Time dependence of the energy of the QDKR with potential
$V_A$ at $\hbar=2\pi+\tilde\hbar$. The red, blue, and green solid bullets
(from left to right) are for, respectively, $\tilde\hbar=0.1$, $10^{-3}$,
and $10^{-5}$. The dotted and the dashed lines indicate the time scaling
$\sim t^2$ and $\sim t^3$, respectively. Two time scales, $t_c$ and $t_s$,
that characterize the $t^2$-$t^3$ crossover and the suppressing of $\sim
t^3$ increasing, are determined by best fitting the corresponding increasing
stage and extrapolating the best fitting lines. (Examples are shown here.)
The results of the measured $t_c$ and $t_s$ for various $\tilde\hbar$ are
presented by circles and triangles, respectively, in (b), as a comparison
with the predicted scaling $t_c\sim1/(K{\tilde\hbar}^{0.5})$ (dashed line)
and $t_s \approx\pi^2 \hbar/(2K\tilde\hbar)$ (dot-dashed line). It can be
seen that both are in good agreement. Panel (c) shows the results of $E(t)$
for $\hbar=2\pi/3+\tilde\hbar$ instead; From left to right, the red, blue,
and green bullets are for $\tilde\hbar=10^{-2}$, $10^{-3}$, and $10^{-4}$,
respectively.}
\label{F:tcts}
\end{figure*}
%%%%%%%%%%%%%%%%%%%%%%%%%%%%%%%%%%%%%%%%%%%%%%%%%%%%%%%%%%%%%%%%%%%%%%%%%%%%%%%%%%figure4

In fact, the initial ballistic spreading of the considered QDKR is a result
of the quantum antiresonance. To show this let us assume $\tilde\hbar=0$
and $\hbar =2\pi$ so that $\exp(\pm i\frac{p^2}{2\hbar})|k\rangle=(-1)^k
|k\rangle$. Taking this into account, we can write the matrix elements of
operator $U$ [Eq.~(\ref{Eq:U})] as
\begin{equation}
\langle j|U|k\rangle =\frac{(-1)^{j+k}}{2\pi}\int_{-\pi}^{\pi}d\theta
e^{i(k-j)\theta} e^{-i\frac{K}{\hbar}[V(\theta)+V(\theta+\pi)]}.
\end{equation}
It follows that for $V_A$, which satisfies $V_A(\theta)=-V_A(\theta+\pi)$,
$U$ reduces to the identity operator~\cite{Dana96}. We have the same
property for $V_B$ but not for $V_C$. Now, if $\tilde\hbar$ is slightly
changed, we find that locally $U$ remains close to the identity, i.e.,
for $|j|, |k| < 1/\sqrt{\tilde\hbar}$,
\begin{equation}
\langle j|U|k\rangle\approx
\left\{\begin{array}{ll}
1-O(\tilde\hbar)  & {\rm for~~} j=k,\\
O(\tilde\hbar^{3/2})  & {\rm for~~} j\ne k.\\
\end{array}\right.
\end{equation}
This gives $\langle j|U^t|k\rangle\approx t \langle j|U|k\rangle$
($j\ne k$) and
\begin{equation}
E(t)=\frac{\hbar^2}{2}\sum_j j^2|\langle j|U^t|k\rangle|^2
\approx D t^2
\end{equation}
for the initial state $|0\rangle$ with
\begin{equation}
D=\frac{\hbar^2}{2}\sum_j j^2|\langle j|U|0\rangle|^2.
\end{equation}
This explains why $E(t)$ increases ballistically at the initial
stage for the QDKR with potential $V_A$ and $V_B$. Numerically,
we can further determine $D$ up to a numerical prefactor, i.e.,
\begin{equation}
D\sim K^2\sqrt{\tilde\hbar}.
\label{Eq:D}
\end{equation}
For the QDKR with potential $V_A$, if we identify its superballistic
spreading process with that of its pseudoclassical system, then by
combining Eq.~(\ref{Eq:t3ratio}) and Eq.~(\ref{Eq:D}), we find
\begin{equation}
t_c\sim\frac{1}{K\sqrt{\tilde\hbar}}.
\label{Eq:tc}
\end{equation}

We put Eq.~(\ref{Eq:tc}) and (\ref{Eq:ts}) into numerical tests and
show the results are in Fig.~\ref{F:tcts}(b). It can be seen that the
simulation results fit them very well. Numerically, $t_c$ and $t_s$
are determined by best fitting $E(t)$ with $\sim t^2$ and $\sim t^3$
respectively in the ballistic and superballistic stage and extrapolating
the two best fitting lines: Their intersection point gives $t_c$ and the
time at which $E(t)$ begins to deviate from the extrapolated best fitting
line over the superballistic stage gives $t_s$ [see Fig.~\ref{F:tcts}(a)].

For other cases in this study where the superballistic spreading is
observed, the numerical factors of $t_c$ and $t_s$ obtained via this
special example are different, but their dependence on $\tilde\hbar$,
i.e., $t_c\sim 1/{\tilde\hbar}^{0.5}$ and $t_s\sim 1/\tilde\hbar$, still
holds, as confirmed by extensive simulations. Hence the scaling relations
given by Eq.~(\ref{Eq:tctsEs}) are expected to be valid generally for the
superballistic spreading in non-KAM QDKR systems.

%%%%%%%%%%%%%%%%%%%%%%%%%%%%%%%%%%%%%%%%%%%%%%%%%%%%%%%%%%%%%%%%%%%%%%%%%%%%%%%%%%figure5
\begin{figure*}[!]
\includegraphics[width=17.5cm]{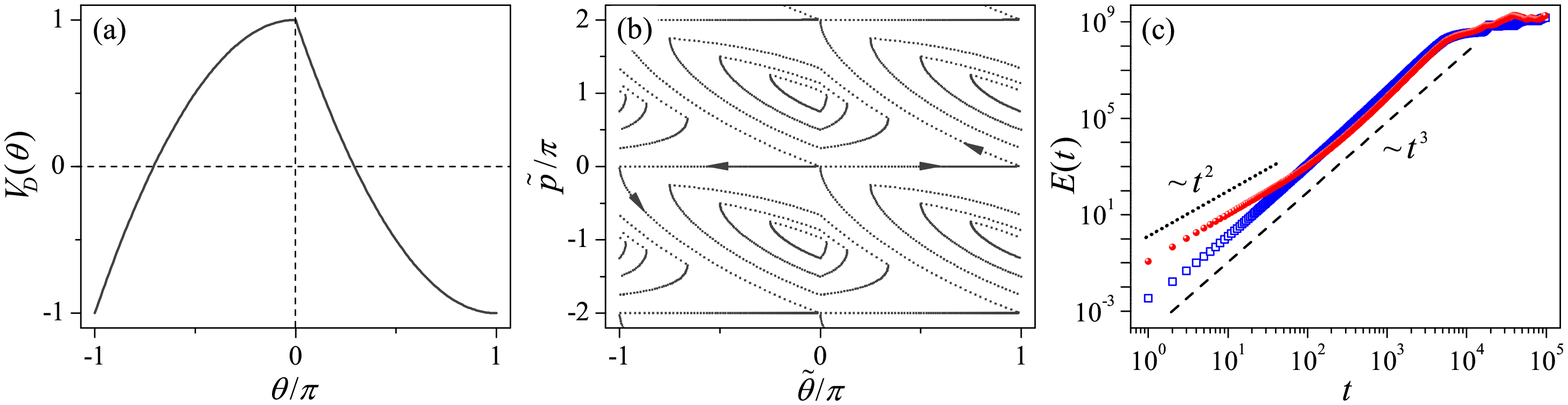}
\caption{The superballistic wavepacket spreading occurs in the QDKR
with a more general non-KAM potential, $V_D$, given by Eq.~(\ref{Eq:VD}).
(a) The potential $V_D$ has the symmetry $V_D(\theta)=-V_D(\theta+\pi)$.
(b) The phase portrait of the corresponding pseudoclassical system.
(c) Time dependence of the energy of the QDKR (red solid bullets) and
its pseudoclassical system (blue open squares). For (b) and (c), the
system parameters are $K=5$, $\hbar=2\pi+\tilde\hbar$ and
$\tilde\hbar=10^{-3}$. }
\label{F:VD}
\end{figure*}
%%%%%%%%%%%%%%%%%%%%%%%%%%%%%%%%%%%%%%%%%%%%%%%%%%%%%%%%%%%%%%%%%%%%%%%%%%%%%%%%%%figure5

\section{VI. Discussions and summary}

In the last two sections we have mainly studied the special case of
$V_A$ and $\hbar=2\pi+\tilde\hbar$ ($\tilde\hbar\ll 1$). In fact the
superballistic wavepacket spreading have been found to generally occur
in the QDKR with other non-KAM potentials of symmetry $V(\theta)=
-V(\theta+\pi)$ and other $\hbar$ values close to $2\pi M/N$. For
example, in Fig.~\ref{F:tcts}(c) we present the simulation results of
$E(t)$ for potential $V_A$ but at $\hbar\approx 2\pi/3$, from which the
characteristics of the superballistic spreading similar to the case of
$\hbar\approx 2\pi$ can be clearly seen. In addition, it has also been
found that $t_c$, $t_s$, and $E_s$ in this case follows
Eq.~(\ref{Eq:tctsEs}) closely (data not shown). Another example is
shown in Fig.~\ref{F:VD} for a more general non-KAM potential:
\begin{equation}
V_D(\theta)=\left\{\begin{array}{ll}
1-2(\theta/\pi)^2 & {\rm ~for~~} -\pi \le \theta < 0;\\
2(\theta/\pi-1)^2-1 & {\rm ~for~~~~~~} 0 \le \theta < \pi.\\
\end{array}\right.
\label{Eq:VD}
\end{equation}
This is a piecewise quadratic potential `randomly' designed with
the only requirement of symmetry $V_D(\theta)=-V_D(\theta+\pi)$.
It has been seen that the superballistic spreading occurs again
with the same characteristics.

It is interesting to notice that the symmetry of potential $V(\theta)
=-V(\theta+\pi)$ plays an important role for both the superballistic
spreading observed in this work and the exponential spreading observed
in Ref.~\cite{Wang11} with $V_S(\theta)=\cos(\theta)$. Indeed, we find
that if we replace $V_S$ with $V(\theta)=\cos(m\theta)$ where $m$ is
instead an even integer, which is a KAM-type potential but loses the
symmetry of $V(\theta)=-V(\theta+\pi)$, then the exponential spreading
no longer happens; rather, it is replaced by the ballistic spreading.
For $V(\theta)=\cos(m\theta)$ with even $m$, we find that the phase space 
structure of the corresponding pseudoclassical system at $\hbar\approx 
2\pi$ is similar to Fig.~\ref{F:PP}(c): The key structure along the 
$\tilde\theta$ axis in the case of $V_S$, i.e., the stable and unstable 
manifolds of the hyperbolic fixed points $(\tilde\theta, \tilde p)=(-\pi,0)$ 
and $(0,0)$ that are essential to the exponential spreading~\cite{Wang11},
vanishes. Therefore, the role the symmetry $V(\theta)=-V(\theta+\pi)$
plays, is to create a special structure along the $\tilde\theta$ axis
which in turn is taken full advantage by the initial conditions around
the $\tilde\theta$ axis. (See the discussion at the end of Sec.~IV.)
In our non-KAM QDKR, the profound difference in quantum dynamics lies
in the phase space structure (of the pseudoclassical system) along the
$\tilde\theta$ axis: Due to the non-KAM nature of the potential, the
hyperbolicity of the fixed points is destroyed and the phase points
on the $\tilde\theta$ axis do not move exponentially.

In summary, we have shown that the superballistic wavepacket spreading
can happen in a class of QDKR systems provides, (i) $\hbar\approx 2\pi M/N$
where $M$ and $N$ are odd and coprime integers; (ii) the system is non-KAM;
and (iii) the potential has the symmetry $V(\theta)=-V(\theta+\pi)$. The
wavepacket spreading will become exponential if the system is KAM-type or
ballistic if the potential loses the required symmetry. Our study evidences
the effectiveness of the pseudoclassical limit theory and the rich dynamics
of the QKR. An interesting question is whether the hyperdiffusive wavepacket
spreading ($\gamma>3$) is possible in the QKR and its variants, which we
leave for future studies.

\section*{Acknowledgements}

This work is supported by the National Natural Science Foundation of China
(Grants No. 11275159, No. 11535011, and No. 11335006).

%%%%%%%%%%%%%%%%%%%%%%%%%%%%%%%%%%%%%%%%%%%%%%%%%%%%%%%%%%%%%%%%%%%%%%%%%%%%%%%%%%%%%%%%%

\end{document}